# A Framework for Smart Homes for Elderly People using Labview®

Ilapakurthy Sriram Vamsi

Computer Engineering, University of California Irvine, Irvine, USA

## Abstract

*This paper deals with home automation systems that are essential for safe and independent living of elderly people. These individuals must be able to perform their Activities of Daily Living (ADLs) without help from caretakers. They must be confident enough that help is just a few minutes with the systems in place. The focus will be on the medical and emergency systems, which are most essential for independent living. Zigbee based Wireless Sensor Networks (WSN) are used to collect the data throughout the home. Smart shirts and smart phones monitor different parameters of the individual and transmit the data to the home network. A PC running LabVIEW is used as the central control unit for the entire house. The house must have a fail proof communication framework for connecting with emergency services, doctors, helpers and relatives. The home network is connected to the Public Switched Telephone Network (PSTN) and internet. SMS, email, prerecorded voice-based telephone calls are appropriately used to report emergencies. A GUI (Graphical User Interface) was built in LabVIEW in order to control and simulate an entire home. Ambient Intelligence is built into the system to make it adaptive and dynamic as far as possible.*

## 1. Introduction

With the advancement of new technologies, the average life expectancy in the world has considerably gone up. The latest developments give freedom and confidence in the lives of elderly people. Children cannot take care of the old people all the time because of various reasons. Elderly people need constant medical supervision and monitoring their condition on a regular basis. Many old people are forced to stay in old age homes, this not only limits their freedom but they also miss the home environment and their family. By having a system which automatically takes care of them they can be happy and can enjoy their old age.

The world's population is not only growing larger, it is also becoming older. The proportion of older persons is increasing at a faster rate than any other age segment. According to the UN Population Division, during the next 45 years, the number of persons in the world aged 60 years or older is expected to almost triple, increasing from 668million people in 2005 to nearly 2.03 billion of older persons live in developing countries; by 2050, that proportion will increase to almost 80 per cent [1]. This shows the need for creating smart homes for elderly people.

In cases of post operations where people need constant monitoring home automation is a cost effective and better option than staying in a hospital. The house can be custom designed based on the disease the person is suffering from. With the automated sensing systems, power consumption of the house is also being greatly reduced. If a person inside a home just falls asleep during the night time forgetting to close the doors and windows, the system can do it by itself by observing the daily schedule of the person and by using adaptive algorithms. The entire house can be controlled remotely through a web browser and the person's details can also be viewed on the site, anyone with valid username and password can access it. The person can have an overall view of the house by just viewing the GUI (Graphical User Interface) on the computer. He can also switch on the lights, fans etc. from the GUI itself reducing physical effort and strain.

This paper describes the implementation of various home automation systems using LabVIEW software. Initially the type of sensors, actuators and the positioning of the sensors in different locations are described. The working and simulation is discussed in the subsequent sections.

## 2. SENSORS

Three types of sensor technologies have demonstrated ability to address the challenges of sensing human activity in a smart home, including wearable devices where sensors are worn by the residents, direct environment components where sensors are distributed in the environment, and infrastructure mediated systems where sensors are installed on an existing home infrastructure. Sensors worn by the residents can be embedded into clothes, eyeglasses, shoes, and wristwatches, or positioned directly on the body. They monitor features that are descriptive of the person's physiological state or movement. Direct environment components typically consists of a set of sensors and an associated sensor network (wired or Zigbee based wireless) to transfer data to a centralized monitoring system where sensor fusion and activity inference take place [2]. Infrastructure mediated systems leverage existing home infrastructure such as the plumbing or electrical system, to mediate the transduction of events [3].

### 2.1. Simple Binary Sensors

One particular type of sensor that is commonly used in smart homes is binary sensor, which simply detects the state of an object or movement with a single digit ‗1'or ‗0'. Various types of binary sensors are being used in smart homes including motion detectors, pressure sensors, and contact switches. Motion detectors and pressure sensors are usually used to detect occupant presence and locations throughout the house [4]. The information about the presence of person at a particular location is an important parameter for the central control system. Contact switches are usually installed on the doors in a smart home such as the front door and doors of cabinets and appliances to provide information on the specific interaction that the occupant performs with objects and appliances. Majeed and Brown classified data logged from contact switches and motion sensors via fuzzy rules into one of six general activities such as sleeping, preparing and eating food, and receiving visitors [5].

### 2.2. Video Cameras

Video cameras are considered high-content sensors which provide rich sources of information both for human observation and for computer interpretation. However, they usually introduce more technical challenges with respect to storage requirements and information extraction, as well as social challenges around privacy when compared with simple binary sensors [6, 7]. They can be programmed in such a way that the video cameras are active only in cases of emergency situations or in case of remote patient monitoring.

## 2.3. Infrastructure Mediated Systems

Compared to installing simple binary sensors throughout an entire house, infrastructure mediate systems require the installation of one or a few sensors along the existing infrastructure, which significantly reduces the cost and complexity of deployment and maintenance. Froehlich et al. developed HydoSense, a customized pressure sensor that could be installed at any accessible location within a home's existing water infrastructure such as an exterior hose bib and utility sink spigot, for activity sensing [8]. Patel et al. developed a powerline noise analyzer that can be plugged into an ordinary wall outlet to detect a variety of electrical events throughout the home. Such data can be used for learning about the daily activities on the person inside the house [9].

## 2.4. Health Monitoring Sensors

BioHarness™ BT [10] is a smart shirt which captures and transmits comprehensive physiological data of the wearer to mobile or to the home network and thus enabling genuine monitoring of human performance and condition in the real world. An accelerometer in the phone or independent module can be used for fall detection of the person and it can directly communicate the information to the central controller. A pneumatic strip installed under the bed linens can be used to detect presence, respiration (normal/abnormal), pulse (low, normal, or high) and movement in the bed, and report four levels of bed restlessness to remote caregivers. Abnormal sleeping patterns can be estimated by comparing the sleeping times with the local database [11].

## 2.5. Some Actuator Mechanisms

Every electronic device is connected to the power supply through a smart switch; the state (on-off) of the device is controlled by the central controller. The doors and windows are capable of automatic closing and opening upon receiving the command. The entire house is fitted with speakers to inform inmates with appropriate information.

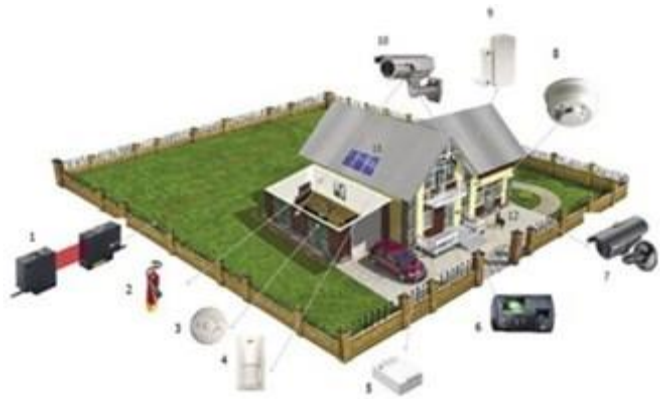

Figure 1. The placement of various sensors in the house

Table 1. Sensors and their functions.

| Id | Name | Description |
| --- | --- | --- |
| 1 | Infrared Beam sensors | It acts an invisible fence around the house. |
| 2 | Fire extinguisher | |
| 3 | Glass break detectors | Detects the sound of breaking glass to raise the alarm. |
| 4 | Indoor Motion detectors | Inform owners about the presence of intruders by detecting human movement. |
| 5 | Gas leak detectors | Detect the presence of dangerous gas |
| 6 | Access control system (card, biometric, retina, palm) | Installed in the main door to prevent unauthorized access. |
| 7 | CCTV cameras | Monitor activities around the house. Infrared cameras can be used for night surveillance. |
| 8 | Smoke detectors | Detects smoke early enough to warn occupants and prevent a fire. |
| 9 | Door Actuators | Automatic opening and closing of doors and windows. |
| 10 | Indoor Video systems | To monitor the patients' activities inside the house. |
| 11 | Solar panels | Charges batteries that allow these gadgets to work. |
| 12 | Canine vigilance | Watch dogs are important to provide security when everything else fails. |

## 3. SIMULATION AND WORKING OF THE SYSTEM

LabVIEW (Laboratory Virtual Instrumentation Engineering Workbench) is a platform and development environment for a visual programming language from National Instruments. It has built in support for hardware integration; it can perform advanced signal analysis and one can build custom user interfaces with it. So, it was chosen to be controller for the smart house.

The sensors throughout the home are either wired to controller or connected wirelessly through the Zigbee network [12]. A base Zigbee receives all the data and relays the data to the Data acquisition (DAQ). A DAQ typically converts analog waveforms into digital values. This data can be accessed by LabVIEW and can know the state of the house at any particular moment.

International Journal on Cybernetics & Informatics (IJCI) Vol. 12, No.4, August 2023

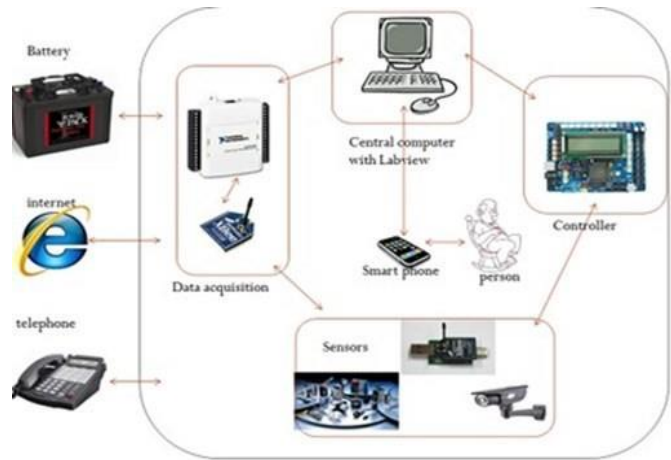

Figure 2. Block diagram of the entire system

### 3.1. Smoke and Gas Alarm

The smoke and gas sensors are polled regularly by the controller; if any abnormal values are detected emergency is declared. All the alarms are raised, nearest fire station is called and all the windows and doors are automatically opened to help the inmates escape the house.

### 3.2. Intrusion Detection

Infrared cameras and motion detectors continuously monitor the compound for any intruders. If anyone is found by the sensors, all the lights are turned on and user is notified of the situation.

### 3.3. Authentication

Before the controller goes into emergency mode by taking input from the sensors, the system gives a certain time for the user to authenticate the action. In case the person finds it to be a false alarm, he can deny the authentication.

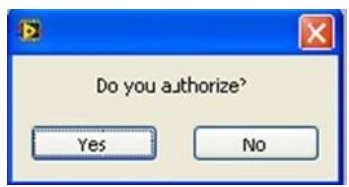

Figure 3. The authentication window in LabVIEW

### 3.4. Web Browser Interface

An application can be published on the web by starting the G server on the LabVIEW. The entire house can be controlled from anywhere in the world. Cameras can be remotely turned on and live video stream can be viewed in the browser. Anyone who needs to access the information will be given usernames and passwords, the rights on what can be controlled remotely will also be established.

International Journal on Cybernetics & Informatics (IJCI) Vol. 12, No.4, August 2023

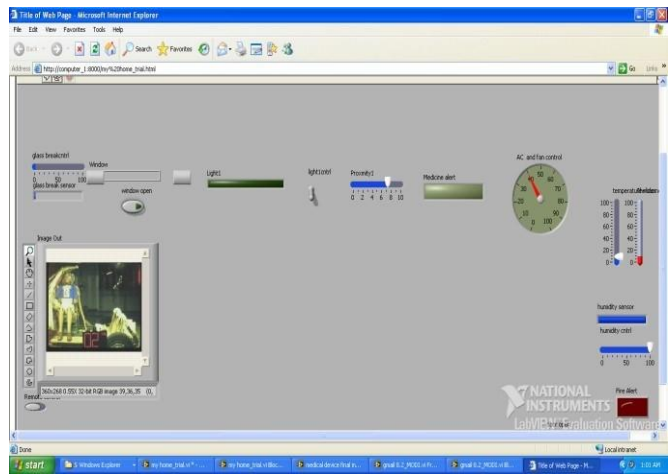

Figure 4. The LabVIEW window showing different controls

### 3.5. Healthcare

The healthcare system would be customized to the needs of a particular individual. The various physiological parameters that are monitored are ECG, heart rate, breathing rate, skin temperature, posture, activity, blood pressure etc. All these signals are processed and compared with the normal limits of the individual; they are also crosschecked with database of various disease models [13]. If any abnormality is found the doctor is called immediately and the patient is given appropriate voice advice to prevent panicking and worsening of the situation. The doctor can see the patient's details by logging on to the patient's computer. He can also have a live video patient examination and can also access the entire recorded parameter database.

### 3.6. Reminders

Old people tend to forget things too often; a proper system should be present for them to have a better life. All the schedules of a person can be loaded in an excel sheet and can be given to the software. The software automatically reminds the person through the speakers about the appointments, medication timings etc.

|   | A | B | C |
|---|---|---|---|
| 1 | Percocet | | |
| 2 | | | |
| 3 | With food? | Yes | |
| 4 | 1-2 tabs every 6 hrs. | | |
| 5 | | | |
| 6 | Day | Time | Dose |
| 7 | Thurs | 8pm | 1 |
| 8 | Friday | 5:00 AM | 1 |
| 9 | | 8:30 AM | 1 |
| 10 | | 2:30 PM | 2 |
| 11 | | 8:30 PM | 2 |
| 12 | Sat | 2:30 AM | 2 |
| 13 | | 8:30 AM | 2 |

Figure 5. A sample excel with medication times

### 3.7. Different Modes of Operation

There are generally four different modes in which the software operates. They are morning mode, night mode, manual mode and emergency mode. In the night mode there is maximum security and everything is looked upon in great detail. If in case the system behaves abnormally due to some bugs or failure, the user can switch to the manual mode. In the emergency mode, all the alarms are raised and system calls for help.

### 3.8. External Communications

The PC is connected to the PSTN [14] and internet to have a fail proof communication network. The PC is interfaced in such a way that the system can make phone calls independently and also send SMS. The system has a database of all the phone numbers and emails of the doctors and caretakers. Appropriate communication channel and prerecorded messages would be sent according to the need and emergency.

### 3.9. Activity Monitoring

The sensory data from throughout the house is analyzed to recognize and monitor basic and instrumental activities of daily living performed by the residents such as bathing, dressing, preparing a meal and taking medication. This allows smart homes to capture patterns reflecting physical and cognitive health conditions and then recognize when the patterns begin to deviate from individualized norms and when atypical behavior that may indicate problems or require intervention [15].

## 4. CONCLUSIONS

In this paper a framework for implementation of smart homes with the use of different sensors and using the LabVIEW framework to integrate them is explained. Different modes of operation of the system are discussed. The web browser interface and telemedicine are the most significant features of the smart house.

## AUTHORS

**Sriram Vamsi Ilapakurthy** is a Senior Software Engineer, he received his Masters Computer Engineering from UC Irvine, California, and a Bachelors in Electronics and Communications Engineering from BITS Pilani, India. His interests include Internet of Things, cyber physical systems and distributed systems.

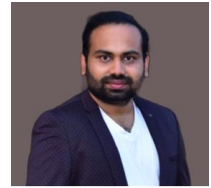